\documentclass{aa}
\usepackage{graphicx}
\usepackage{txfonts}
\usepackage{natbib}
\bibpunct{(}{)}{;}{a}{}{,}
\usepackage{longtable}
\usepackage{lscape}
\usepackage{times}
\usepackage{textcomp}

\begin{document}

\title{Discovery of a new supernova remnant G150.3+4.5}
\subtitle{}
\author{X. Y.~Gao and J. L.~Han}
\titlerunning{G150.3+4.5}
\authorrunning{Gao \& Han}

\offprints{bearwards@gmail.com}

\institute{National Astronomical Observatories, CAS, Jia-20 Datun
  Road, Chaoyang District, Beijing 100012, PR China}

\date{Received 06 May 2014; accepted 04 June 2014}

\abstract
{Large-scale radio continuum surveys have good potential for
  discovering new Galactic supernova remnants (SNRs). Surveys of the
  Galactic plane are often limited in the Galactic latitude of $|b|
  \sim 5\degr$. SNRs at high latitudes, such as the Cygnus Loop or
  CTA~1, cannot be detected by surveys in such limited latitudes.}
{Using the available Urumqi $\lambda$6\ cm Galactic plane survey data,
  together with the maps from the extended ongoing $\lambda$6\ cm
  medium latitude survey, we wish to discover new SNRs in a large sky
  area.}
{We searched for shell-like structures and calculated radio spectra
  using the Urumqi $\lambda$6\ cm, Effelsberg $\lambda$11\ cm, and
  $\lambda$21\ cm survey data. Radio polarized emission and evidence
  in other wavelengths are also examined for the characteristics of
  SNRs.}
{We discover an enclosed oval-shaped object G150.3+4.5 in the
  $\lambda$6\ cm survey map. It is about 2$\fdg$5 wide and 3$\degr$
  high. Parts of the shell structures can be identified well in the
  $\lambda$11\ cm, $\lambda$21\ cm, and $\lambda$73.5\ cm
  observations. The Effelsberg $\lambda$21\ cm total intensity image
  resembles most of the structures of G150.3+4.5 seen at
  $\lambda$6\ cm, but the loop is not closed in the northwest. High
  resolution images at $\lambda$21\ cm and $\lambda$73.5\ cm from the
  Canadian Galactic Plane Survey confirm the extended emission from
  the eastern and western shells of G150.3+4.5. We calculated the
  radio continuum spectral indices of the eastern and western shells,
  which are $\beta \sim -2.4$ and $\beta \sim -2.7$ between
  $\lambda$6\ cm and $\lambda$21\ cm, respectively. The shell-like
  structures and their non-thermal nature strongly suggest that
  G150.3+4.5 is a shell-type SNR. For other objects in the field of
  view, G151.4+3.0 and G151.2+2.6, we confirm that the shell-like
  structure G151.4+3.0 very likely has a SNR origin, while the
  circular-shaped G151.2+2.6 is an \ion{H}{II} region with a flat
  radio spectrum, associated with optical filamentary structure,
  H$\alpha$, and infrared emission.}
{skip}

\keywords{Radio continuum: ISM -- ISM: supernova remnants -- ISM:
  individual objects: G150.3+4.5}

\maketitle
\section{Introduction}

There is a large discrepancy between the theoretically predicted
numbers of the Galactic supernova remnants (SNRs)
\citep[e.g.,][]{Berkhuijsen84, Li91, Tammann94} and the observational
detections \citep{Green09, Ferrand12}, which is often attributed to
the limitations of the current observations on the angular resolutions
to detect distant small SNRs, the sensitivities to reveal faint SNRs,
and the insufficient sky coverage for the completeness of
detections. In addition, strong confusion from the diffuse Galactic
emission makes discoveries of SNRs in the inner Galaxy even more
difficult.

Previously, the big steps in the discovery of the Galactic SNRs relied
on the radio surveys carried out by powerful telescopes, e.g.,
\citet{Shaver70} and \citet{Clark75} on the 408~MHz and 5~GHz surveys
by the Parkes and Molonglo telescopes. \citet{Reich88c} found many
SNRs using the $\lambda$21\ cm and $\lambda$11\ cm surveys made by the
Effelsberg 100-m telescope. With the Molonglo telescope again,
\citet{Whiteoak96} identified SNRs with the 843~MHz survey data. The
most recent significant increase in the number of the Galactic SNRs
was made by \citet{Brogan06}, using the high angular resolution of the
VLA to detect 35 new SNRs in the inner Galaxy. There are also some
individual discoveries, although not many. The Sino-German
$\lambda$6\ cm polarisation survey of the Galactic
plane\footnote{http://zmtt.bao.ac.cn/6cm/} \citep{Sun07, Gao10,
  Sun11a, Xiao11} conducted by the Urumqi 25-m radio telescope
provides a good hunting ground for discovering new Galactic
SNRs. \citet{Gao11y} identified two new faint SNRs G25.1$-$2.3 and
G178.2$-$4.2. Supplemented by the Urumqi $\lambda$6\ cm data,
\citet{Foster13} identified another two new SNRs G152.4$-$2.1 and
G190.9$-$2.2 from the Canadian Galactic Plane Survey (CGPS) data. With
the Giant Meterwave Radio Telescope, \citet{Roy13} have recently
discovered a young SNR G354.4+0.0 in the central region of the
Galaxy. The CGPS has also contributed several other detections,
e.g. G85.4+0.7 and G85.9$-$0.6 by \citet{Kothes01}, G107.5$-$1.5 by
\citet{Kothes03}, G96.0+2.0 and G113.0+0.2 by \citet{Kothes05},
G108.2$-$0.6 by \citet{Tian07}, and G141.2+5.0 by \citet{Kothes14}.
However, most of the radio continuum surveys of the Galactic plane are
limited in a latitude range, e.g. $|b| \leqslant 5\degr$. This is not
sufficient to discover and study the SNRs off the plane, such as the
Cygnus Loop (G74.0$-$8.5) \citep[e.g.,][]{Harris60, Sun06} and the
CTA~1 (G119.5+10.2) \citep[e.g.,][]{Walsh55, Sun11c}.

Following the Urumqi $\lambda$6\ cm plane survey, a survey extending
the latitude range from $b = 5\degr$ to $b = 20\degr$ has been carried
out since 2012 (Gao et al. in prep.). It currently collects the data
of $\lambda$6\ cm total intensity and linear polarization from the
longitude range from $\ell = 90\degr$ to 160$\degr$.  Here we report
the discovery of a new SNR G150.3+4.5 by combing the new data with the
$\lambda$6\ cm plane survey data. We introduce the data in
Sect.~2. The radio continuum emission, the spectrum, and the evidence
collected in other bands are discussed in Sect.~3. We summarize our
work in Sect.~4.

\begin{figure*}[!tbhp]
\centering
\begin{minipage}[bth]{0.32\textwidth}
\centering
\includegraphics[angle=-90, width=5.5cm]{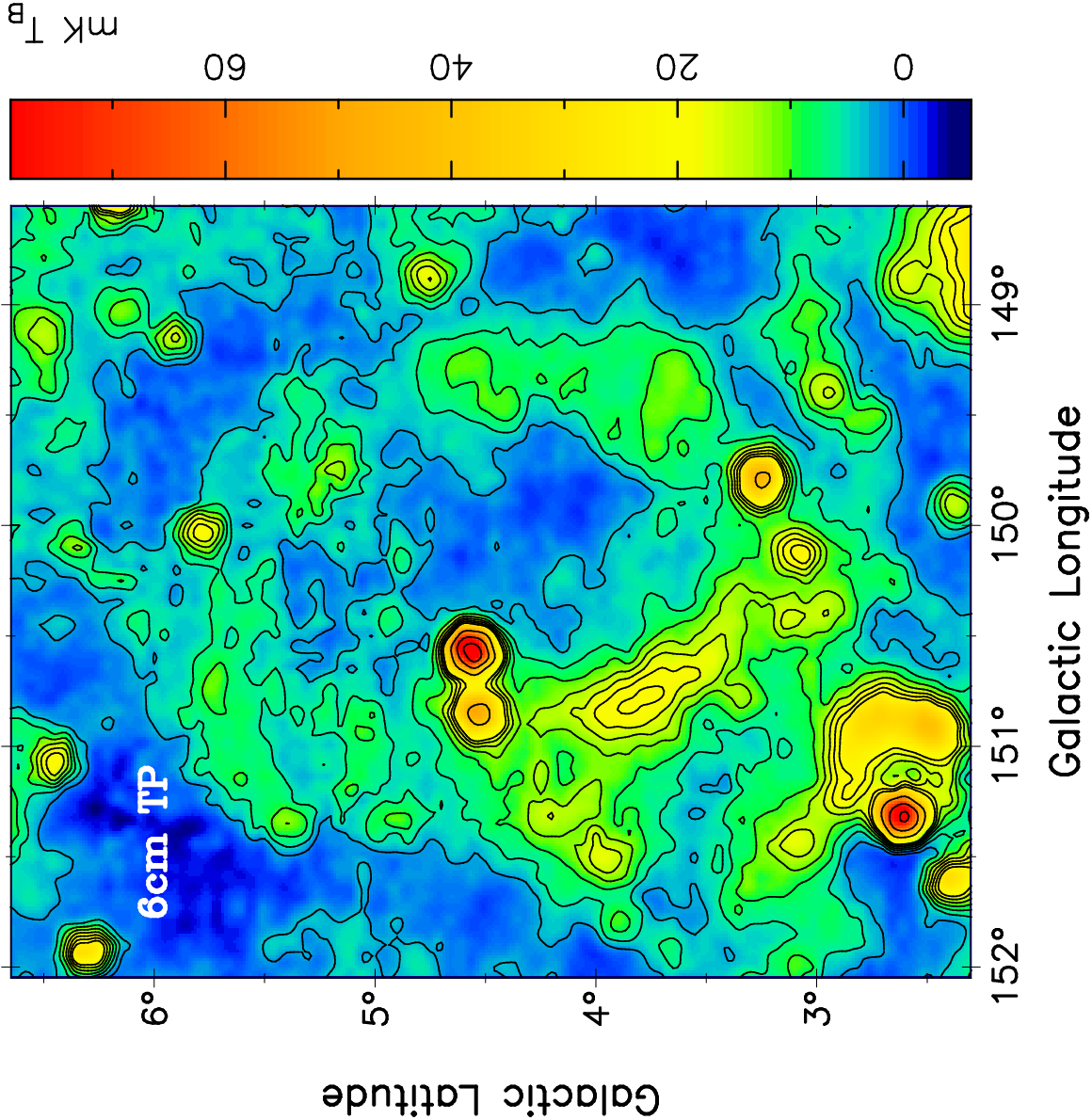}
\end{minipage}
\begin{minipage}[bth]{0.32\textwidth}
\centering
\includegraphics[angle=-90, width=5.5cm]{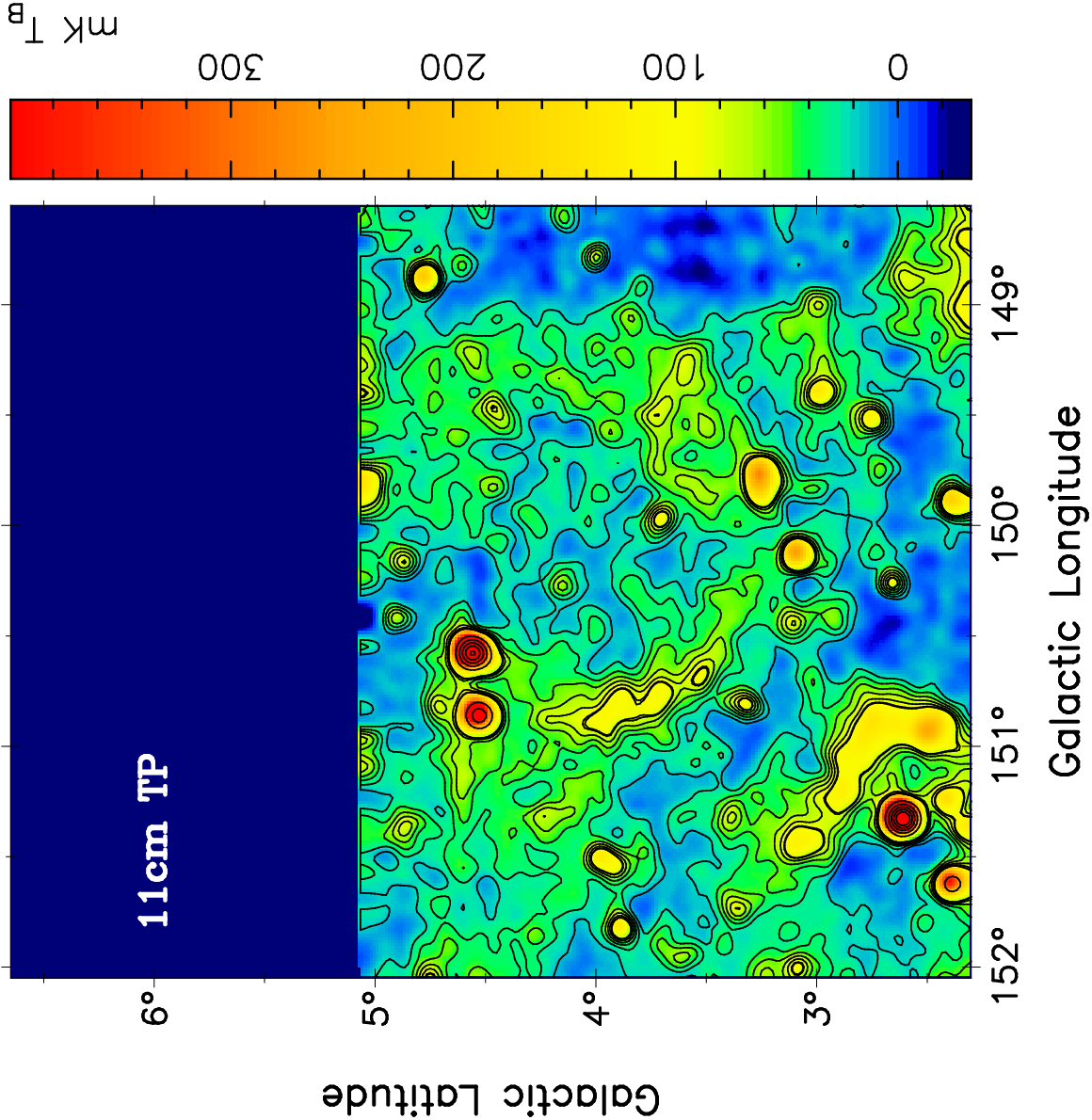}
\end{minipage}
\begin{minipage}[bth]{0.32\textwidth}
\centering
\includegraphics[angle=-90, width=5.5cm]{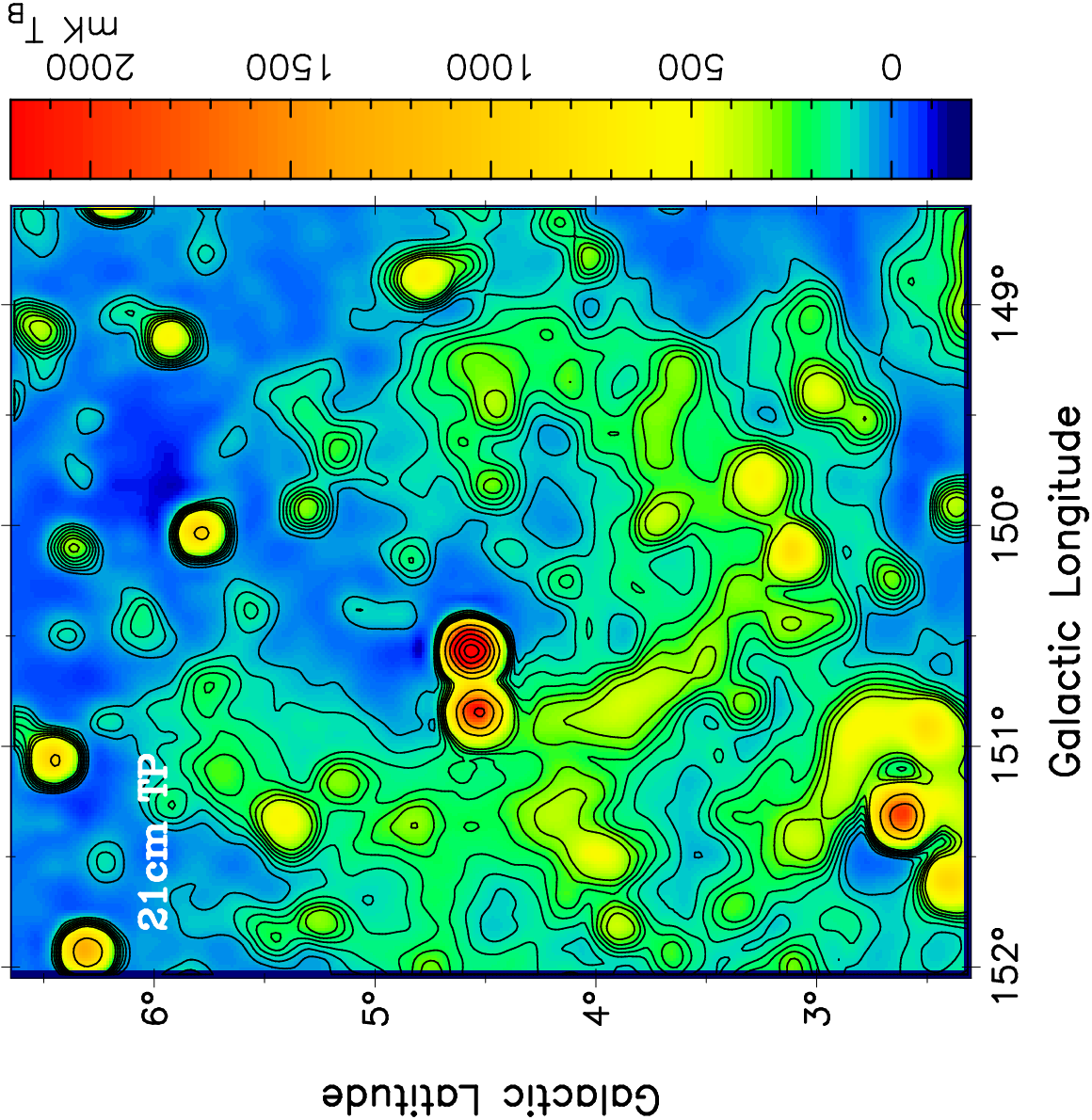}
\end{minipage}
\begin{minipage}[bth]{0.32\textwidth}
\centering
\vspace{0.5cm}
\includegraphics[angle=-90, width=5.5cm]{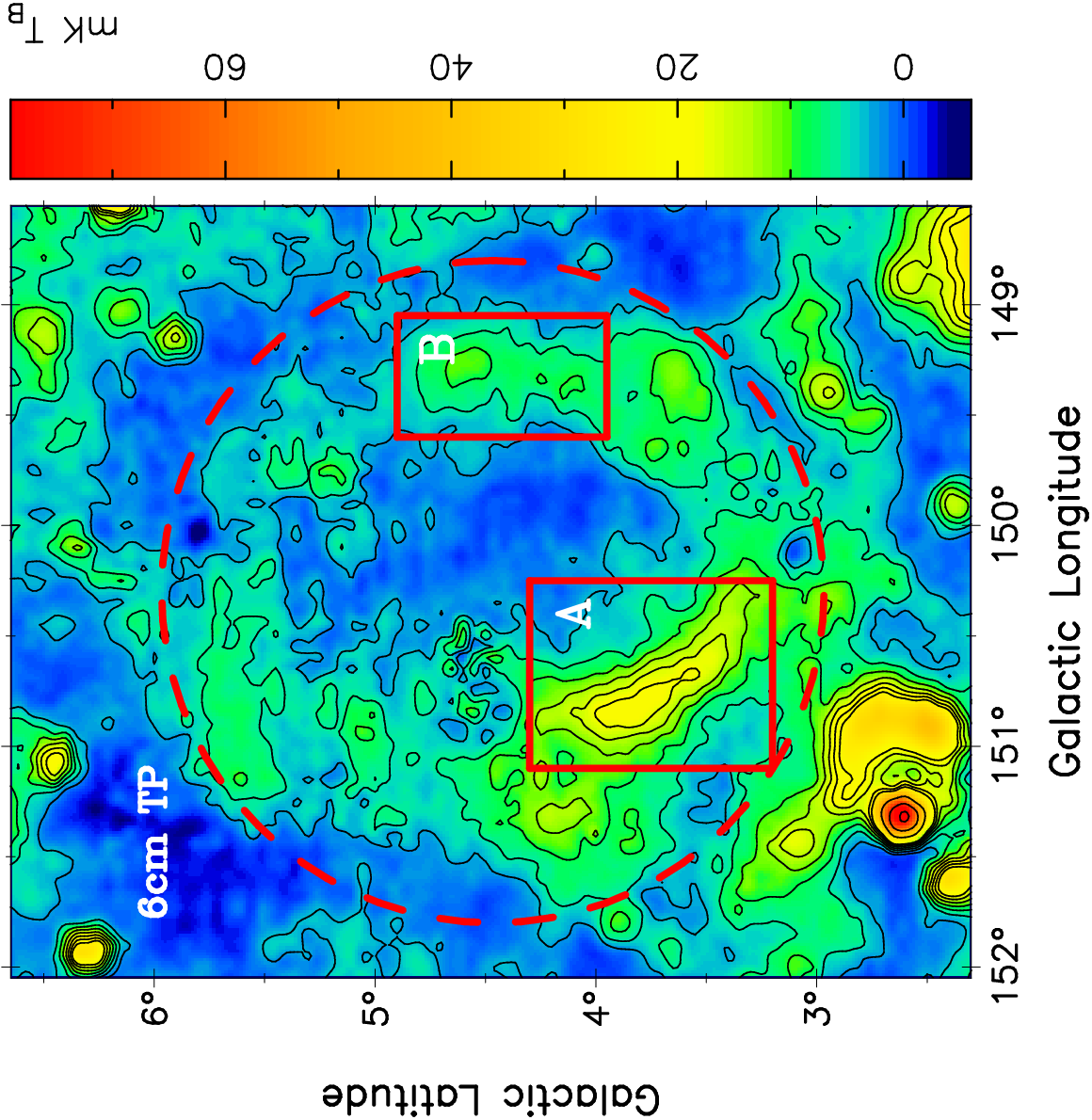}
\end{minipage}
\begin{minipage}[bth]{0.32\textwidth}
\centering
\vspace{0.5cm}
\includegraphics[angle=-90, width=5.5cm]{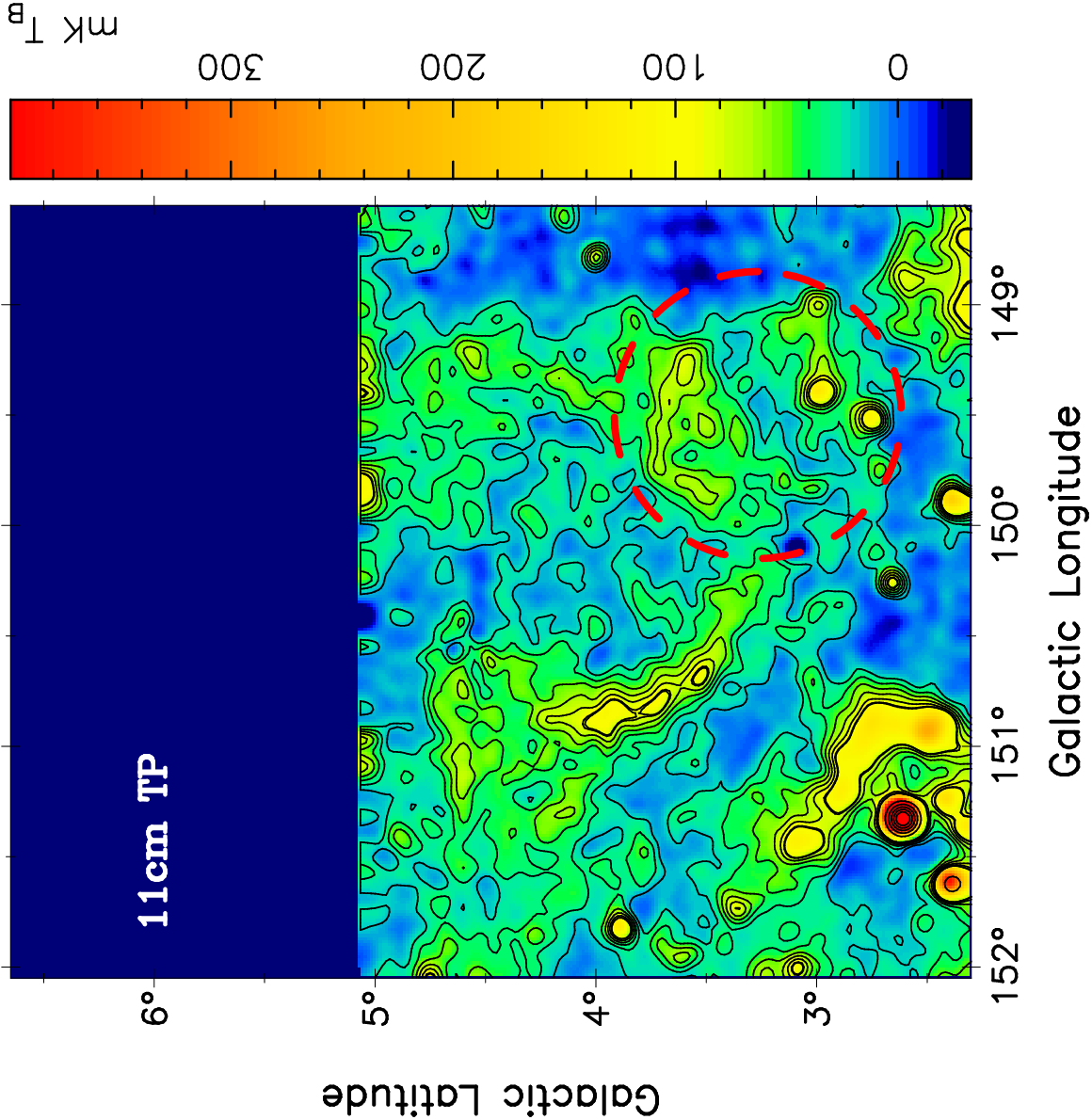}
\end{minipage}
\begin{minipage}[bth]{0.32\textwidth}
\centering
\vspace{0.5cm}
\includegraphics[angle=-90, width=5.5cm]{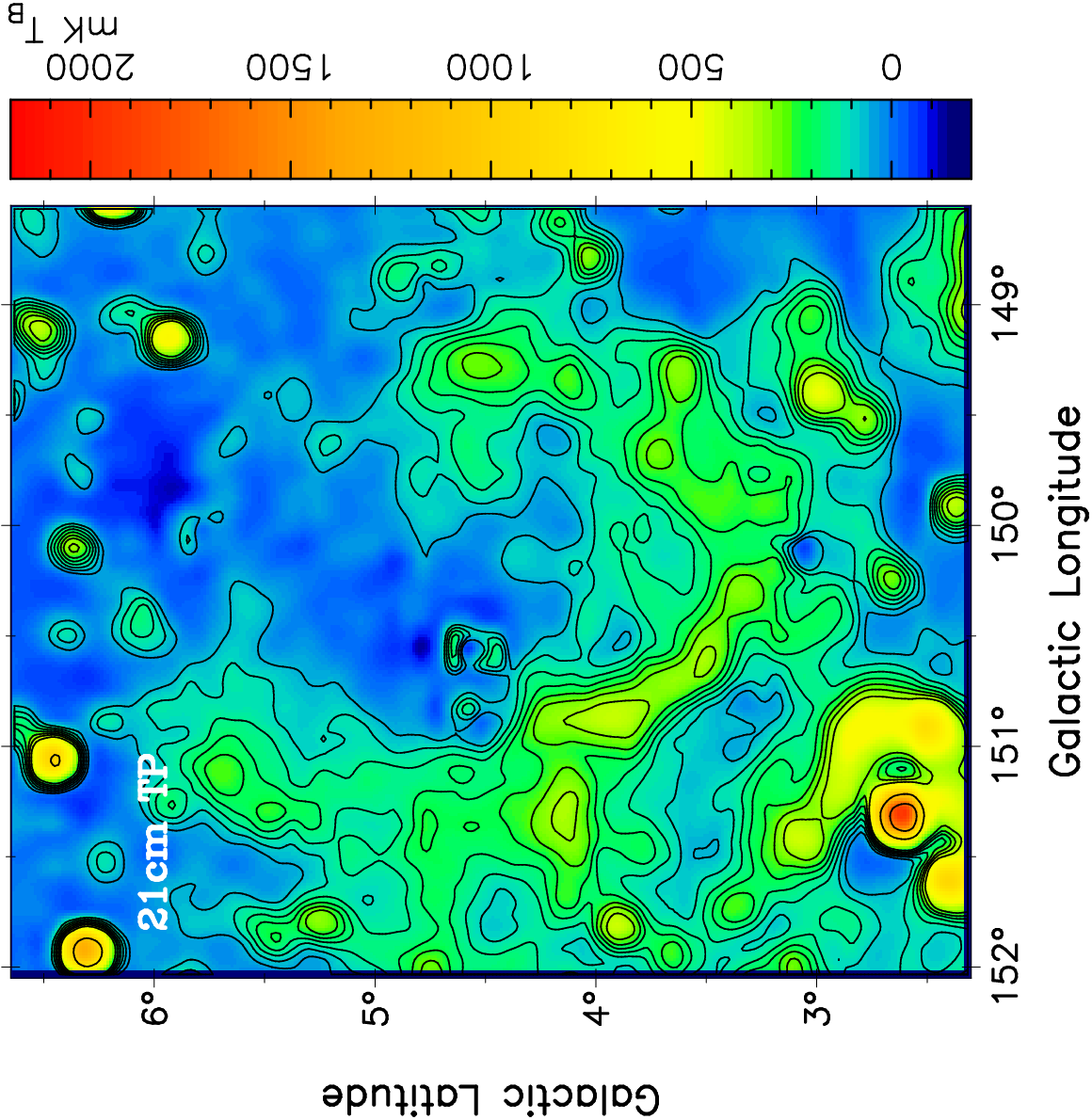}
\end{minipage}
\begin{minipage}[bth]{0.32\textwidth}
\centering
\vspace{0.5cm}
\includegraphics[angle=-90, width=5.5cm]{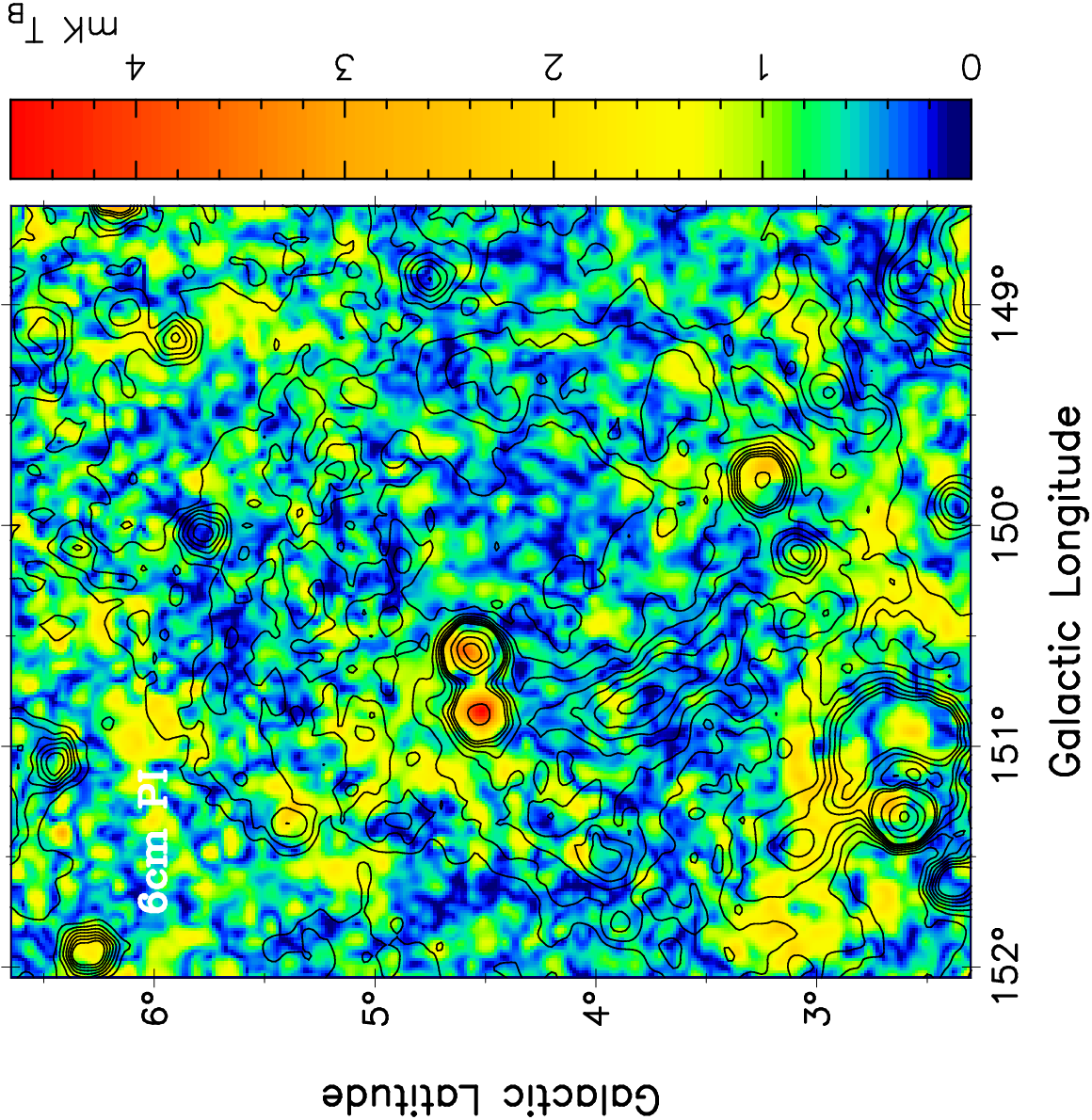}
\end{minipage}
\begin{minipage}[bth]{0.32\textwidth}
\centering
\vspace{0.5cm}
\includegraphics[angle=-90, width=5.5cm]{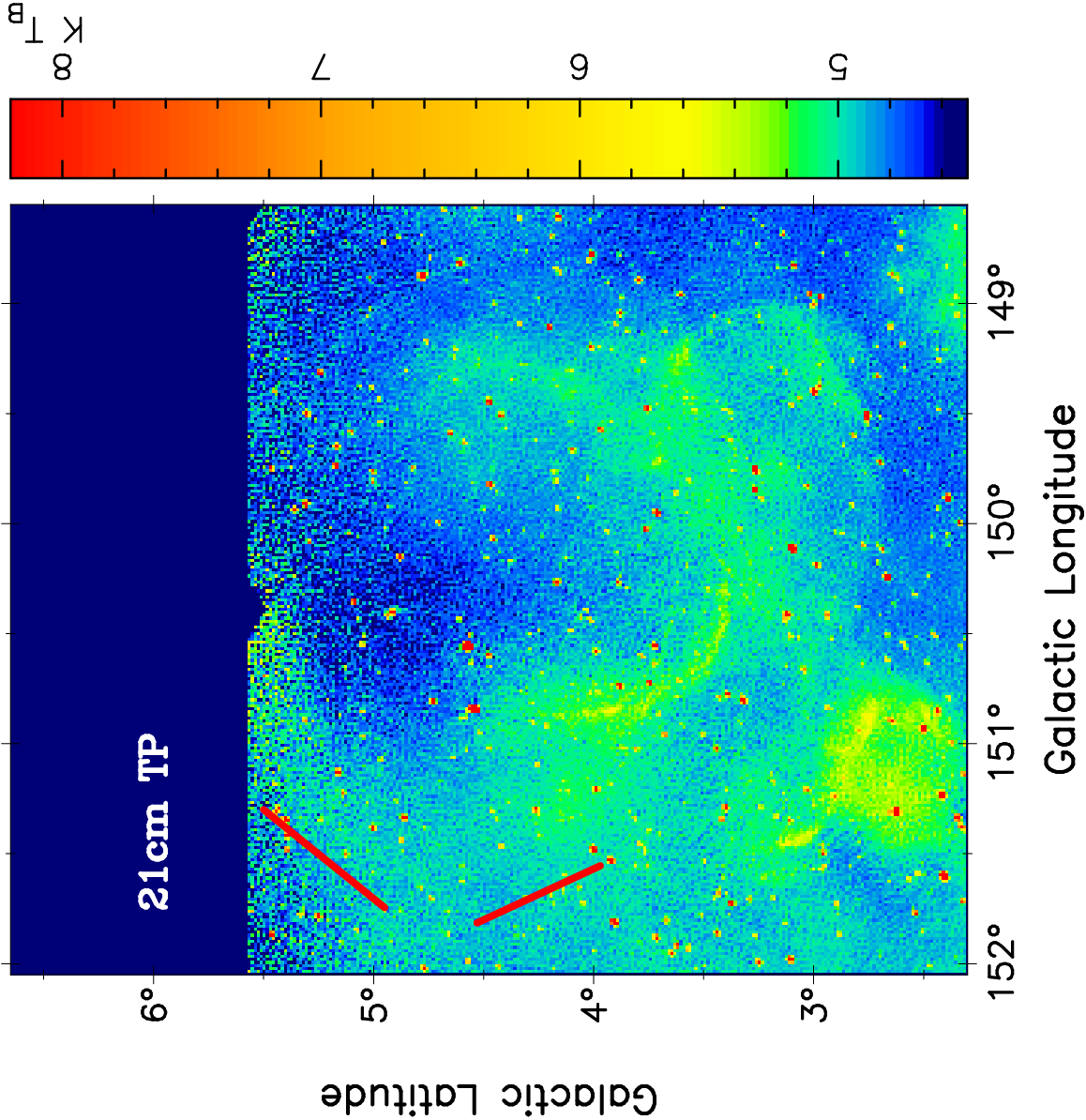}
\end{minipage}
\begin{minipage}[bth]{0.32\textwidth}
\centering
\vspace{0.5cm}
\includegraphics[angle=-90, width=5.5cm]{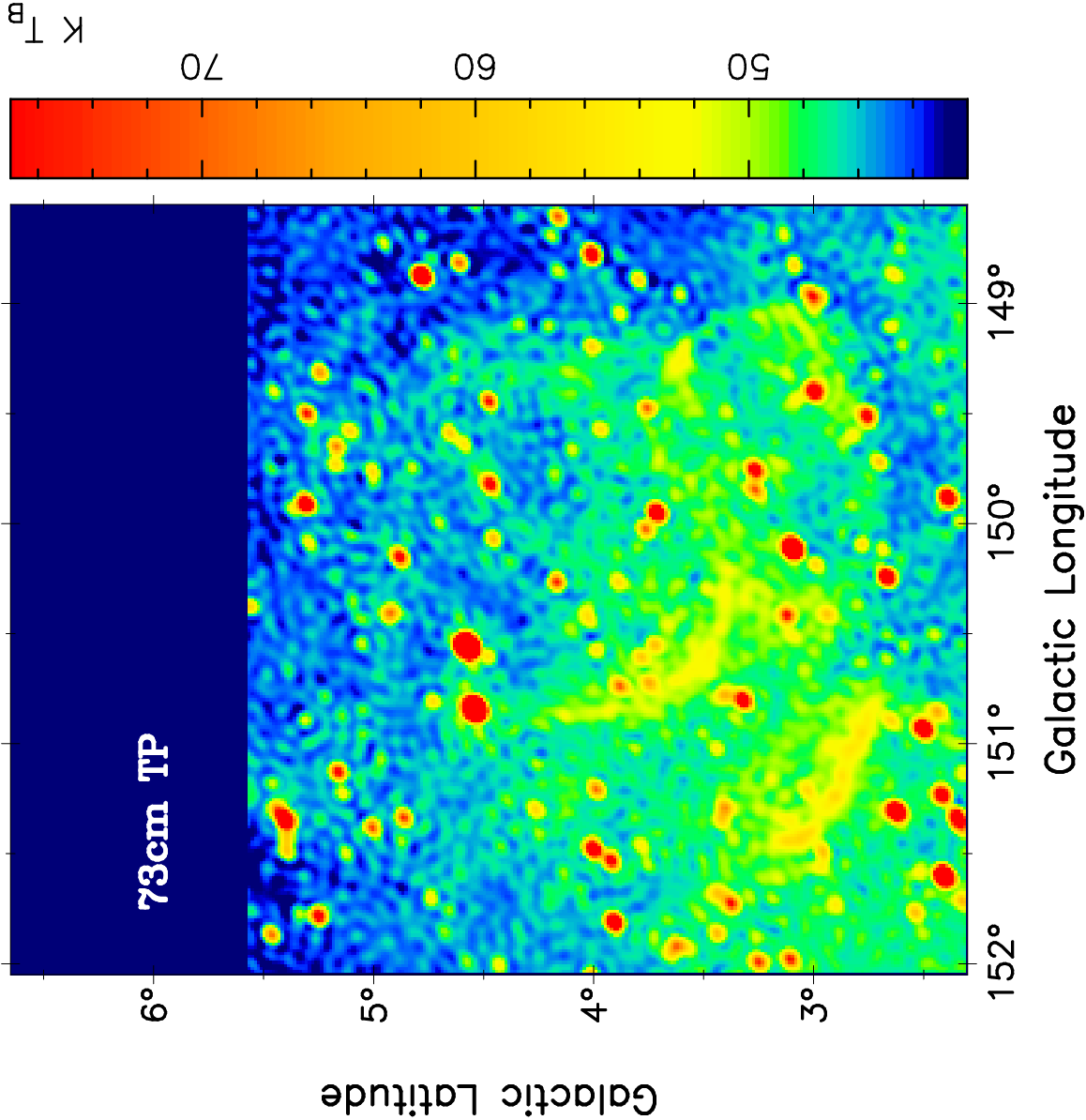}
\end{minipage}
\caption{Radio continuum images of the new SNR G150.3+4.5. {\it Top
    panels:} total intensity of G150.3+4.5 obtained from Urumqi
  $\lambda$6\ cm ({\it left}), Effelsberg $\lambda$11\ cm ({\it
    central}), and Effelsberg $\lambda$21\ cm ({\it right})
  observations. The contours run at 3.0 + (n-1) $\times$ 3.0~mK (n =1,
  2, ...6) and 21.0 + (n-7) $\times$ 21.0~mK (n = 7, 8..) for the
  $\lambda$6\ cm image, at 20.0 + (n-1) $\times$ 13.4~mK (n =1, 2,
  ...6) and 85.0 + (n-7) $\times$ 212.0~mK (n = 7, 8..) for the
  $\lambda$11\ cm image, at 60.0 + (n-1) $\times$ 40.0~mK (n =1, 2,
  ...6) and 320.0 + (n-7) $\times$ 600.0~mK (n = 7, 8..) for the
  $\lambda$21\ cm image. {\it Middle panels:} the same images as {\it
    upper panels}, but point-like sources are removed within the
  central 3$\degr$ field as indicated by the circle in the left
  panel. {\it Bottom panels:} images for the $\lambda$6\ cm
  polarization intensity ({\it left}), the CGPS $\lambda$21\ cm total
  intensity ({\it central}) and $\lambda$73.5\ cm total intensity
  ({\it right}). The contours on the $\lambda$6\ cm polarisation image
  is the same as for the $\lambda$6\ cm total intensity image shown in
  the {\it upper panels}. The angular resolutions for observations of
  the Urumqi $\lambda$6\ cm, Effelsberg $\lambda$21\ cm, CGPS
  $\lambda$21\ cm, and $\lambda$73.5\ cm images are 9$\farcm$5,
  9$\farcm$4, $60'' \times 49''$, and $3\farcm5 \times 2\farcm8$,
  respectively. The Effelsberg $\lambda$11\ cm image was convolved to
  the resolution of 6$\arcmin$ to increase the signal-to-noise
  ratio. The rectangles A and B as indicated in the {\it middle left
    panel} are the areas for TT plots in Fig.~\ref{tt}. The dashed
  circle in the {\it middle central panel} indicates the newly
  discovered SNR G149.5+3.2 by \citet{Gerbrandt14}, which overlaps
  with the lower part of the western shell of G150.3+4.5. The two
  straight lines shown in the CGPS $\lambda$21\ cm total intensity
  image indicate the emission, which does not seem to be connected to
  the eastern shell of G150.3+4.5.}
\label{tp}
\end{figure*}
%

\section{Data}

\begin{figure}[tbp]
\centering
\includegraphics[angle=-90, width=0.23\textwidth]{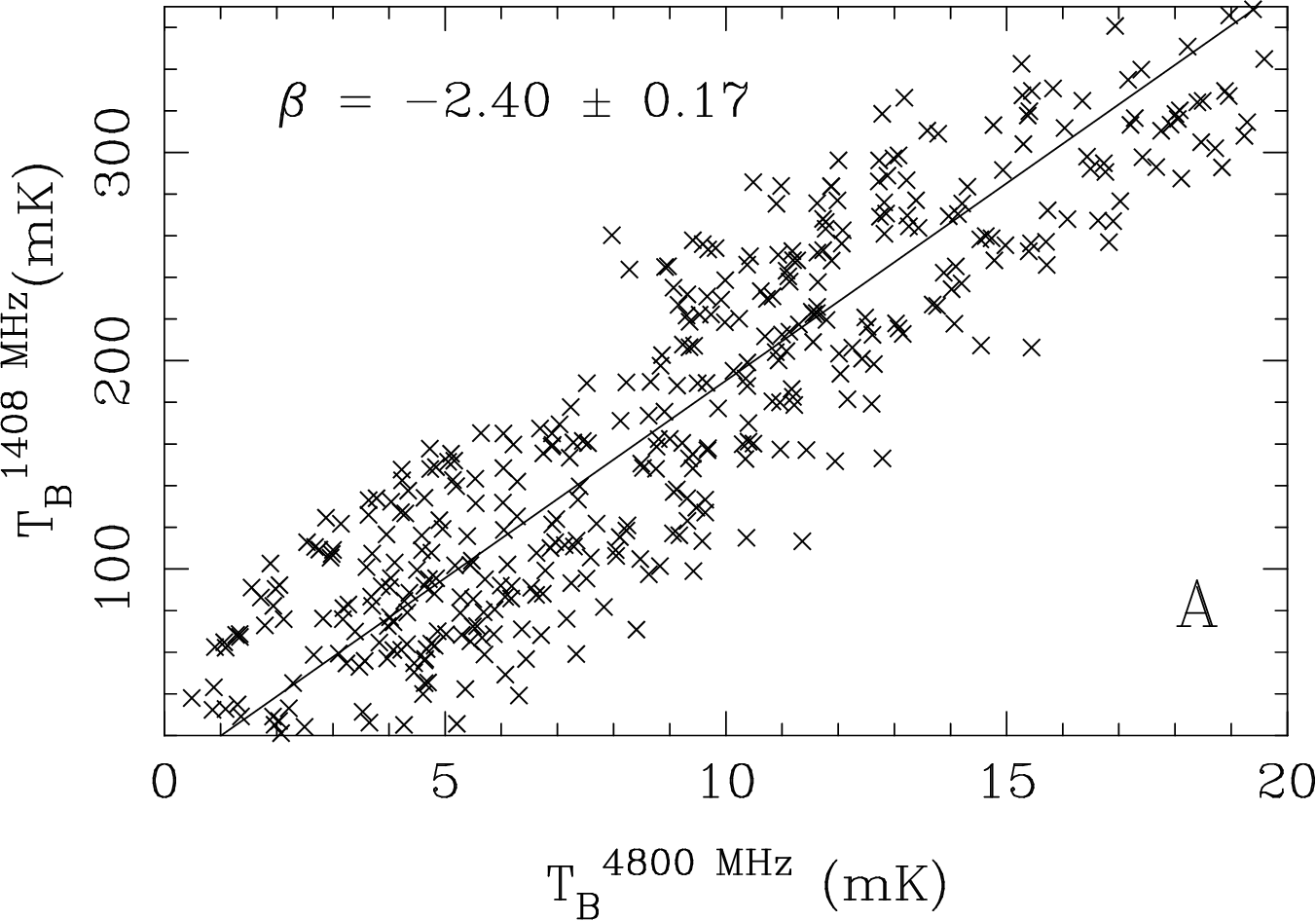}
\includegraphics[angle=-90, width=0.23\textwidth]{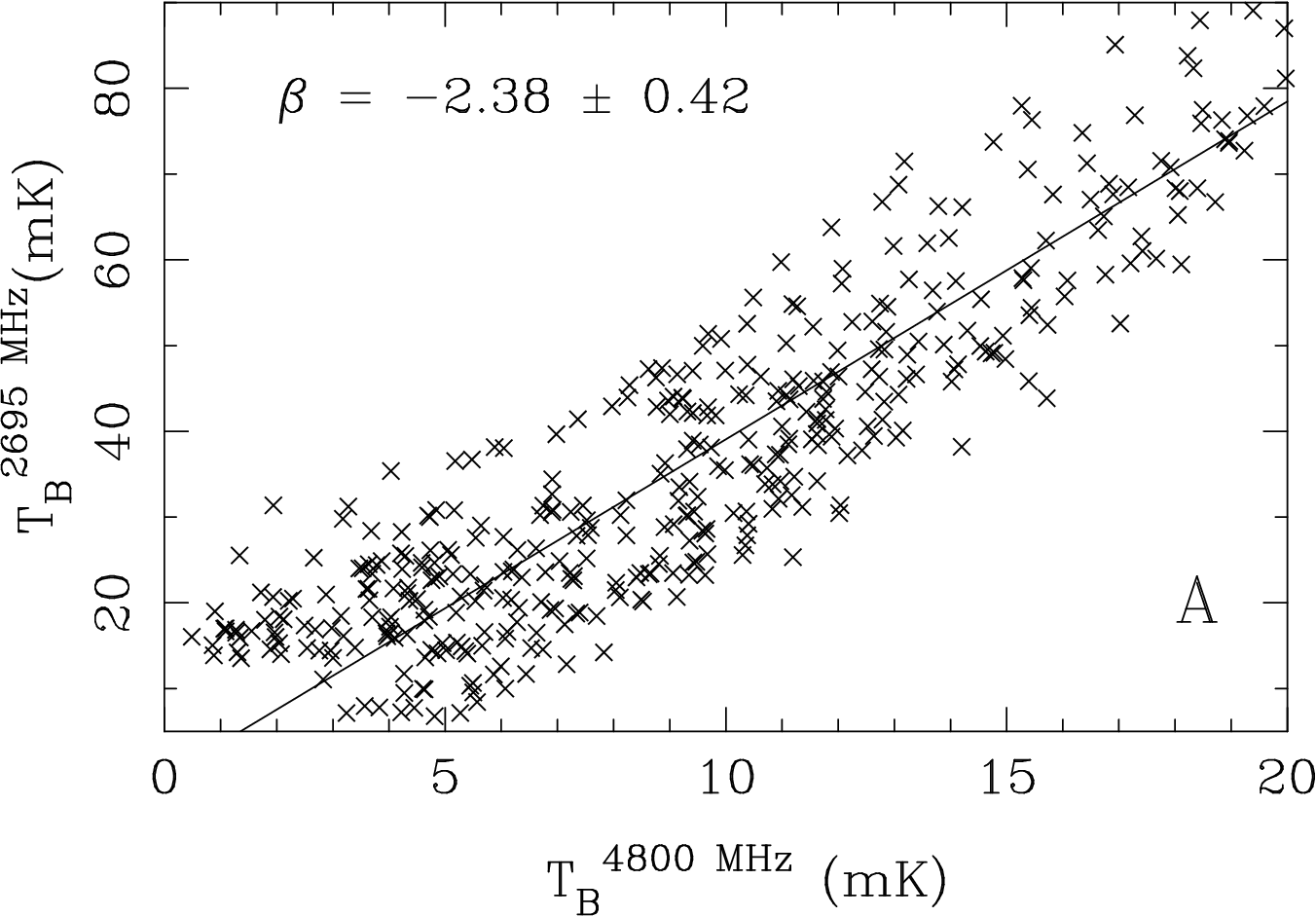}\\
\vspace{0.5cm}
\includegraphics[angle=-90, width=0.23\textwidth]{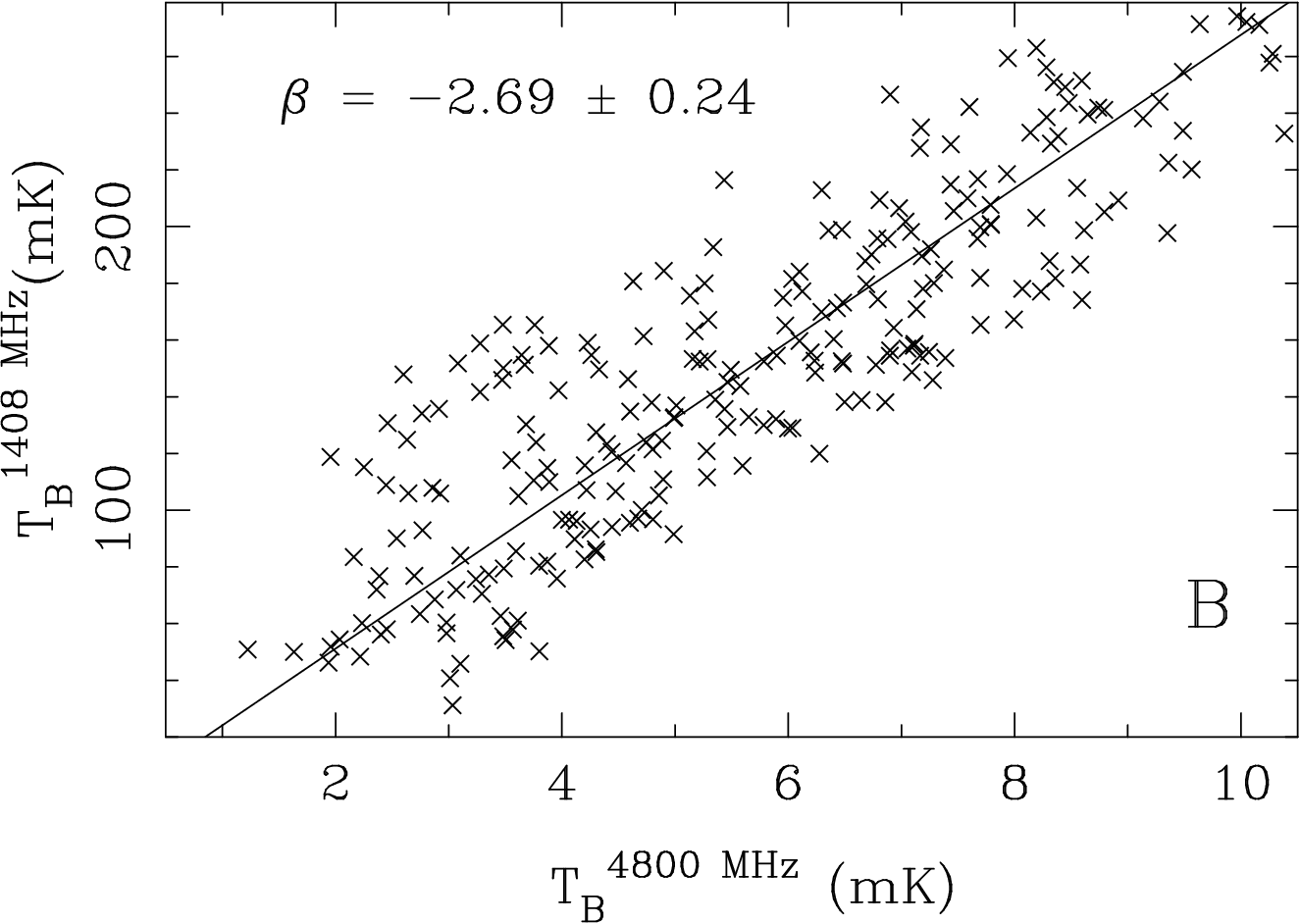}
\includegraphics[angle=-90, width=0.23\textwidth]{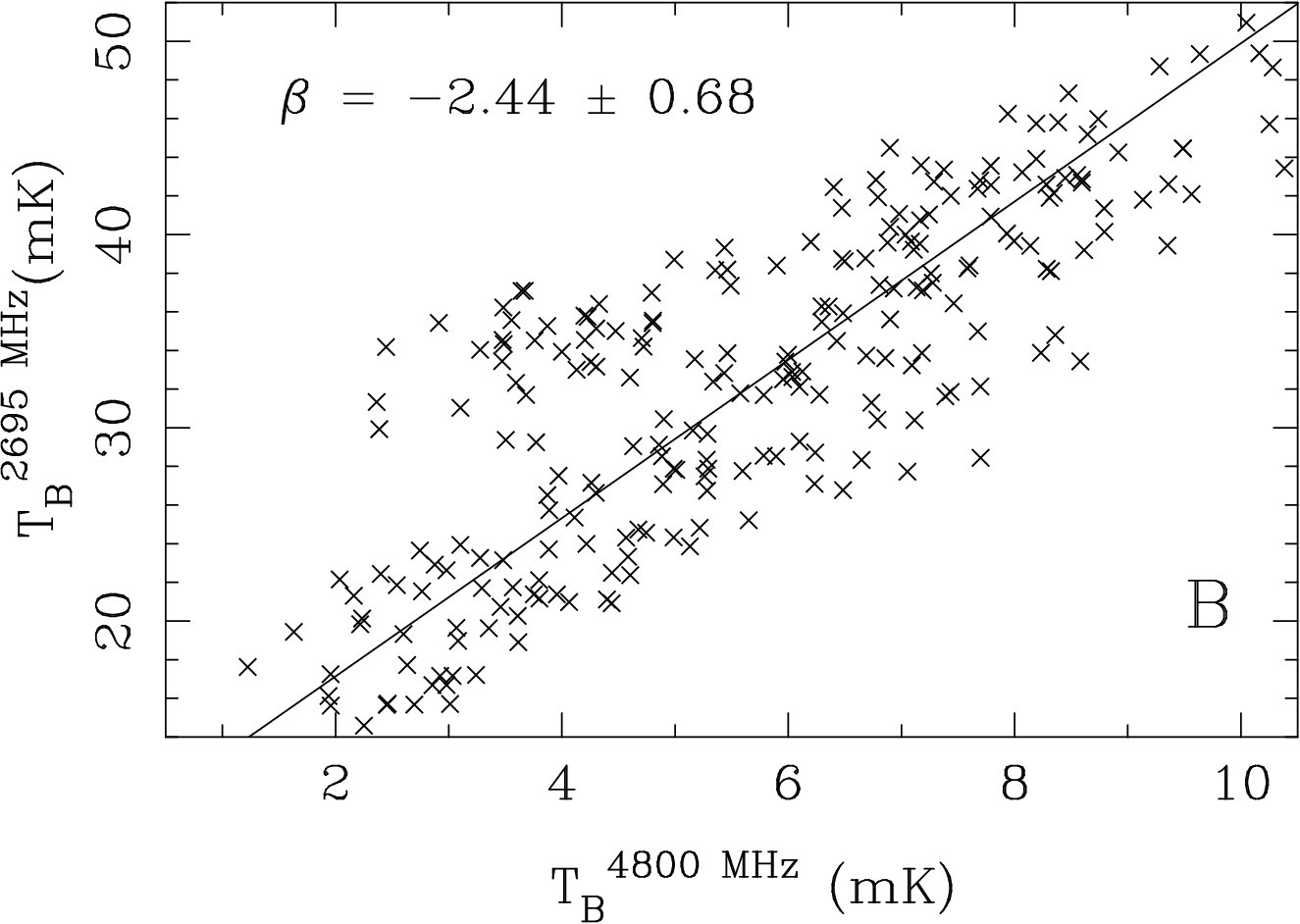}
\caption{TT plots between $\lambda$6\ cm and $\lambda$21\ cm and
  between the $\lambda$6\ cm and $\lambda$11\ cm for the eastern
  shell, i.e., region A ({\it upper panels}), and western shell, i.e.,
  region B ({\it lower panels}) of G150.3+4.5, as indicated by the
  rectangles in Fig.~\ref{tp}.}
\label{tt}
\end{figure}

\subsection{Urumqi $\lambda$6\ cm data}

The $\lambda$6\ cm total intensity $I$ and the linear polarization $U$
and $Q$ data used in this paper come from a portion of the Sino-German
$\lambda$6\ cm polarization survey of the Galactic plane \citep{Gao10}
and the extended survey with the latitude range from $b = 5\degr$ to
$b =20\degr$. Observations of both surveys were made with the same
$\lambda$6\ cm system mounted on the Urumqi 25-m radio telescope,
Xinjiang Astronomical Observatories, Chinese Academy of Sciences. The
observation strategy, data reduction, and calibration follow the same
procedures, which have been described in detail in \citet{Sun07} and
\citet{Gao10}. The angular resolution of the $\lambda$6\ cm map is
9$\farcm$5. The sensitivity is about 1.0\ mK~T$_{b}$ for total
intensity and about 0.3\ mK~T$_{b}$ for polarization.

\subsection{Effelbserg $\lambda$11\ cm data}

The $\lambda$11\ cm total intensity map is a part of the radio
continuum survey of the Galactic plane made by the Effelsberg 100-m
radio telescope \citep{Fuerst90}. We extracted the data from the
survey sampler of the Max-Planck-Institut f{\"u}r
Radioastronomie\footnote{http://www3.mpifr-bonn.mpg.de/survey.html}. The
map has an angular resolution of 4$\farcm$3. We measured 6.7~mK
T$_{b}$ as the r.m.s. (1$\sigma$) for the $\lambda$11\ cm total
intensity data.

\subsection{Effelsberg $\lambda$21\ cm data }

The $\lambda$21\ cm radio continuum data was observed by the
Effelsberg telescope for two parts: the data above the latitude of $b
= 4\degr$ come from the Effelsberg Medium Latitude Survey (EMLS)
\citep{Uyaniker98, Reich04}, while the data below $b = 4\degr$ come
from the Effelsberg $\lambda$21\ cm radio continuum plane survey
\citep{Reich97}. The angular resolution is about 9$\farcm$4,
comparable to that of the $\lambda$6\ cm data. The r.m.s of the map is
about 20.0~mK T$_{b}$.

\subsection{Canadian $\lambda$21\ cm \& $\lambda$73\ cm data}

Both of the Canadian $\lambda$21\ cm (1420~MHz) and $\lambda$73.5\ cm
(408~MHz) data are from the Canadian Galactic Plane Survey (CGPS)
conducted by the synthesis telescopes of the Dominion Radio
Astronomical Observatory (DRAO) \citep{Taylor03, Landecker10}. The
data were extracted from the the Canadian astronomy data
center\footnote{http://www1.cadc-ccda.hia-iha.nrc-cnrc.gc.ca/cgps/query.html}. The
$\lambda$21\ cm (1420~MHz) data has an angular resolution of $60''
\times 49''$ in the area of $\ell = 150\fdg3, b = 4.5$, while the
$\lambda$73.5\ cm (408~MHz) data has a beam of 3$\farcm5 \times
2\farcm$8. In this paper, we use the CGPS maps for qualitative study
and for showing fine structures on small scales.

\section{Results}

We present the Urumqi $\lambda$6\ cm, the Effelsberg $\lambda$11\ cm,
and the Effelsberg $\lambda$21\ cm total intensity images of a $3\fdg5
\times 4\fdg35$ region around the target in the top panels of
Fig.~\ref{tp}. The loop structure seen from the $\lambda$6\ cm image
is named G150.3+4.5, according to its geometric center. To study the
faint and extended emission of this object, prominent point-like
sources (S$_{\rm 1.4GHz}$ $>$ 20~mJy) within the central field of
3$\degr$ are subtracted based on the NVSS source catalog
\citep{Condon98}. For point-like sources that have determined radio
flux density spectral indices ($S_{\nu} = \nu^{\alpha}$)
\citep{Vollmer05}, we simply extrapolate the flux densities from
1.4~GHz to the other two frequencies. For sources with unknown
spectral indices, a spectral index of $\alpha = -0.9$ is used for
extrapolation. We show the total intensity images in the middle panels
of Fig.~\ref{tp} for the $\lambda$6\ cm, $\lambda$11\ cm and the
Effelsberg $\lambda$21\ cm bands after the point-like sources are
removed. The $\lambda$6\ cm polarization image and the high resolution
total intensity images of the DRAO $\lambda$21\ cm and
$\lambda$73.5\ cm are shown in the bottom panels of Fig.~\ref{tp}. The
currently available radio continuum observations of the Effelsberg
$\lambda$11\ cm data ends up at $b = 5\degr$, while the DRAO
$\lambda$21\ cm and $\lambda$73.5\ cm data extend to about $b =
5\fdg5$. The angular resolutions of the $\lambda$6\ cm, the Effelsberg
$\lambda$21\ cm, the DRAO $\lambda$21\ cm, and the DRAO
$\lambda$73.5\ cm images are 9$\farcm$5, 9$\farcm$4, $60'' \times
49''$ and $3\farcm5 \times 2\farcm8$, respectively. To increase the
signal-to-noise ratio, the Effelsberg $\lambda$11\ cm total intensity
image was convolved to an angular resolution of 6$\arcmin$.

\subsection{Total intensity and the spectral index}

\begin{figure*}[tbh]
\centering
\includegraphics[angle=-90, width=0.3\textwidth]{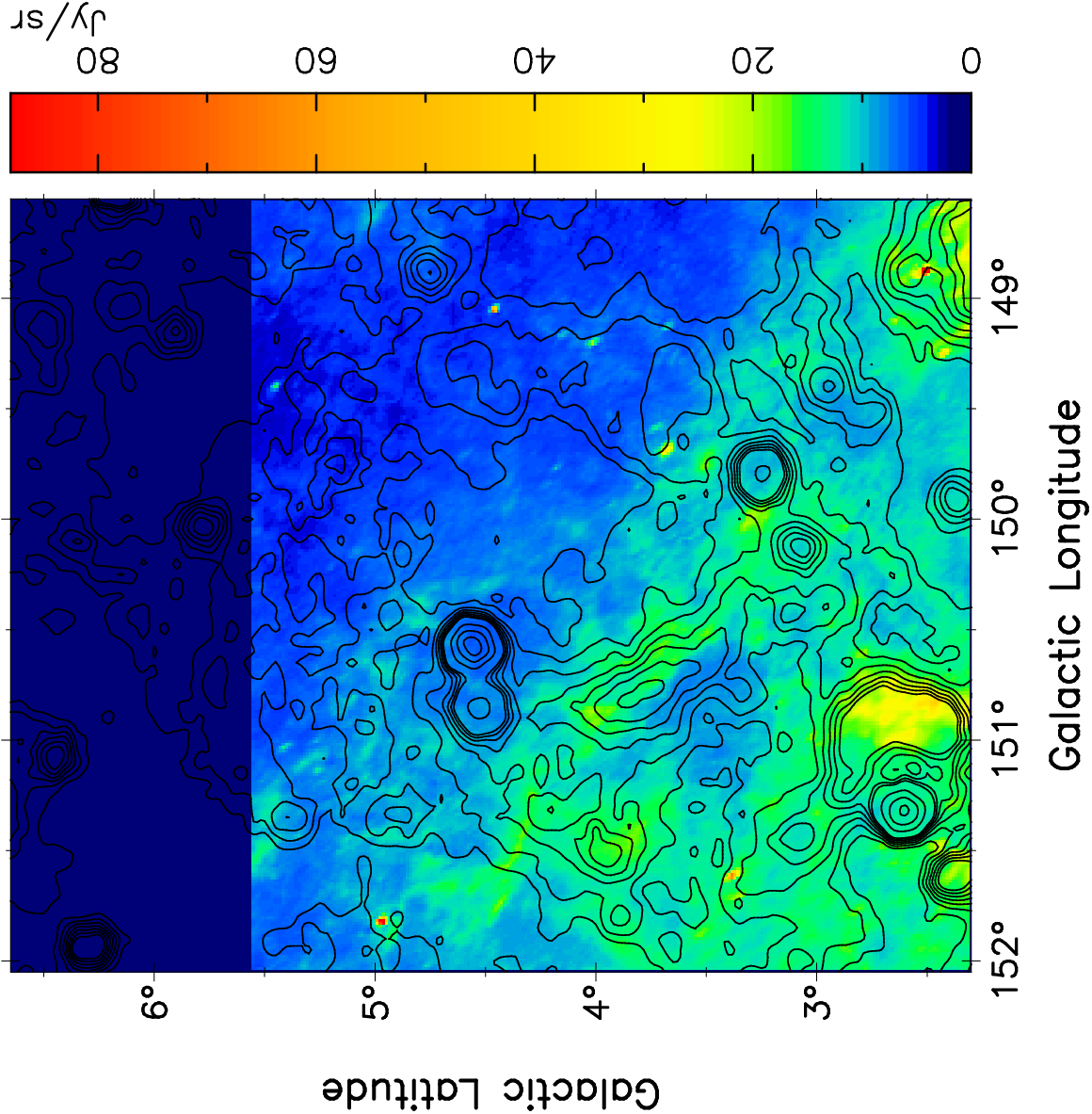}
\includegraphics[angle=-90, width=0.3\textwidth]{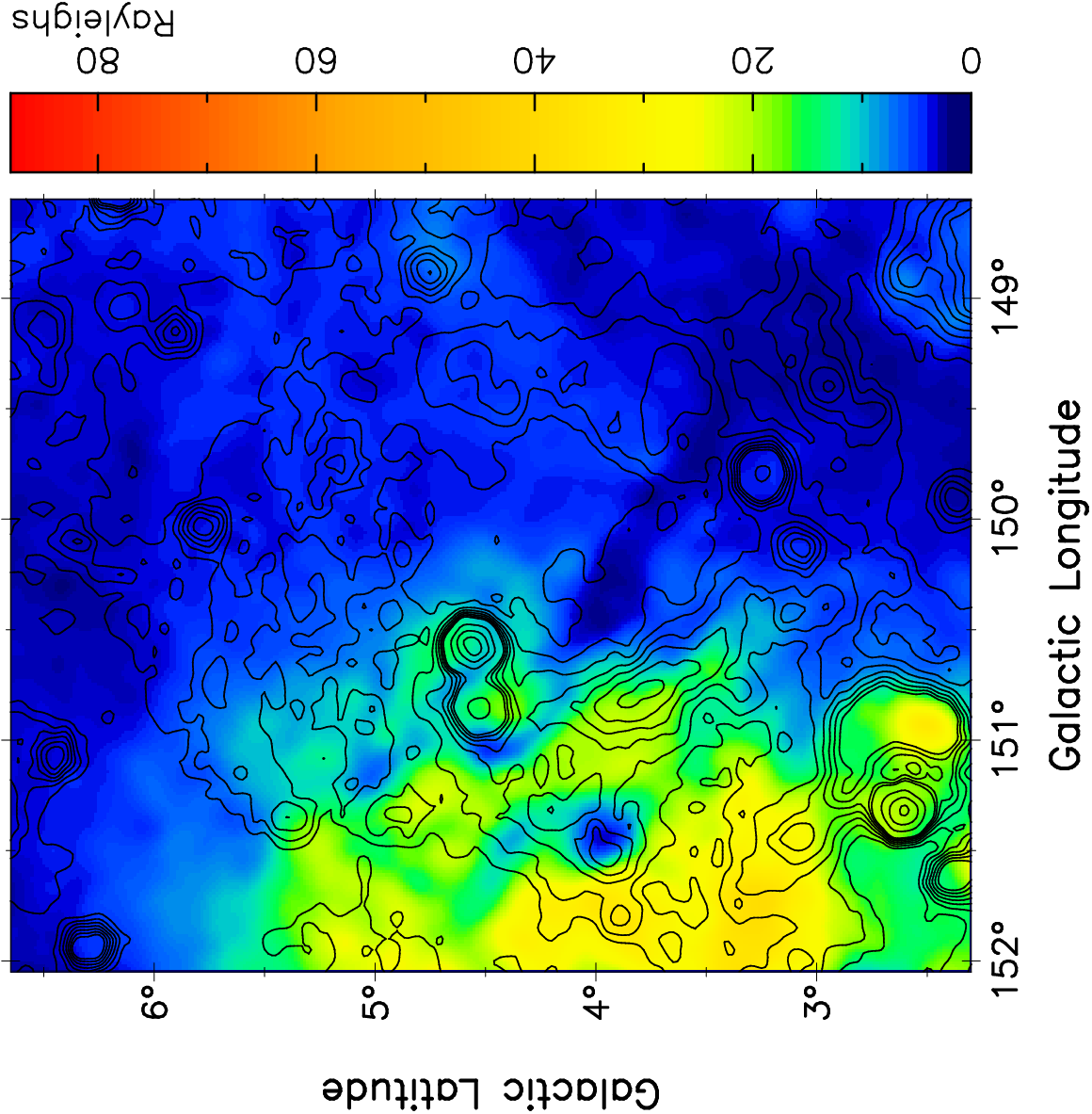}
\includegraphics[angle=-90, width=0.3\textwidth]{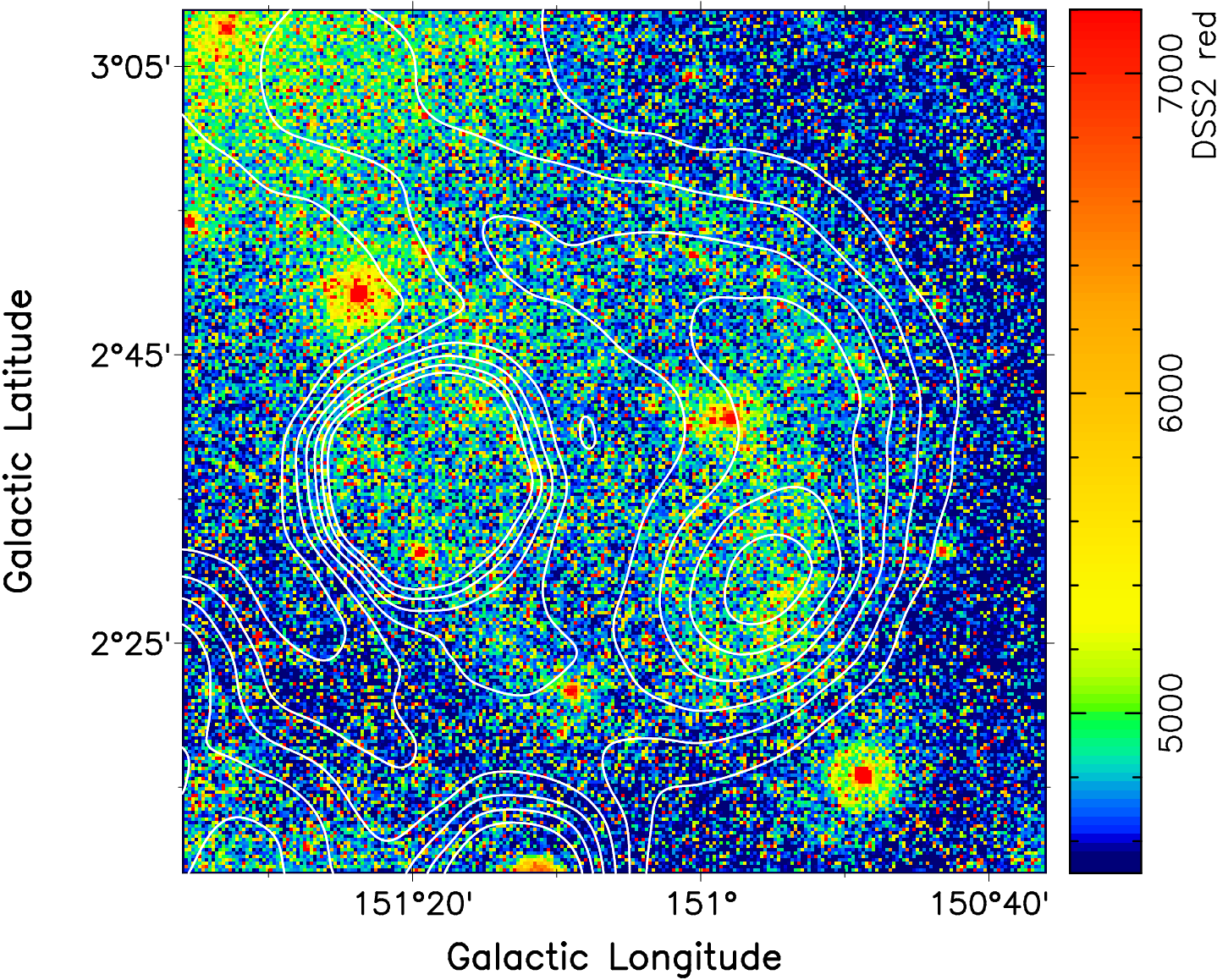}
\caption{60$\mu$m infrared ({\it left panel}) and H$\alpha$ emission
  ({\it middle panel}) in the area of G150.3+4.5. The contours are the
  same for the $\lambda$6\ cm image shown in the {\it upper panel} of
  Fig.~\ref{tp}. {\it Right panel:} optical emission from the DSS2 red
  image in the field of G151.2+2.6. $\lambda$6\ cm total intensity
  contours run at 6.0~mK T$_{b}$ in steps of 6.0~mK T$_{b}$.}
\label{other}
\end{figure*}

The big loop G150.3+4.5 is clearly seen in the Urumqi $\lambda$6\ cm
and the Effelsberg $\lambda$21\ cm images, extending to a higher
latitude of more than $b = 6\degr$. It shows an enclosed oval shape at
$\lambda$6\ cm, which is about 2$\fdg$5 wide and 3$\degr$ high. Three
major shells of G150.3+4.5 can be identified in the $\lambda$6\ cm
image. The most prominent one is found in the east, curving to the
center of the lower south, while a fainter shell can be identified in
the west, also extending to the lower south. A much fainter and
fragmented shell is found in the upper north. From the $\lambda$6\ cm
map after subtracting the point-like sources, we see the emission in
that region is just around 3$\sigma$ ($\sim$ 3.0~mK T$_{b}$) detection
above the background.

Most of the structures seen at $\lambda$6\ cm also appear in the
Effelsberg $\lambda$21\ cm total intensity map, but the loop is not
closed, with a gap in the northwest. This part is detected at
$\lambda$6\ cm to be about 3.0~mK T$_{b}$, but below 3$\sigma_{21cm}$
= 60~mK T$_{b}$ at $\lambda$21\ cm (see Fig.~\ref{tp}), which
indicates that the brightness spectral index of this gap region is
shallower than $\beta = -2.46$ ($S_{\nu} = \nu^{\alpha}$, $\beta =
\alpha - 2$).

The $\lambda$11\ cm data cover only the lower part of G150.3+4.5, but
reveals more detailed structures with a 6$\arcmin$ angular
resolution. The new SNR G149.5+3.2, as indicated in Fig.~\ref{tp} has
just been discovered by \citet{Gerbrandt14}. It overlaps with the
lower part of the western shell of G150.3+4.5.

The DRAO $\lambda$21\ cm image has the sharpest resolution. The
eastern shell of G150.3+4.5 can be traced clearly. The overlap of the
lower part of the western shell with the new SNR G149.5+3.2
\citep{Gerbrandt14} is clearly seen. The upper part of the western
shell, free of the contamination of G149.5+3.2, can be identified
above $b = 4\degr$.  The fragmented northern shell of G150.3+4.5 seen
at $\lambda$6\ cm cannot be identified. From the $\sim 1\arcmin$
resolution map, there is no continuous emission of the prominent
eastern shell pointing to the north. Faint extended emission goes
wider from $\ell = 151\fdg3, b = 4\fdg1$ to $\ell = 151\fdg6, b
=4\fdg5$ and then turns to $\ell = 150\fdg9, b = 5\fdg4$ (as indicated
in Fig.~\ref{tp}). As for the Effelsberg $\lambda$21\ cm map, the DRAO
$\lambda$21\ cm observations do not detect the emission in the
northwest of the loop, either.

The DRAO $\lambda$73.5\ cm image revealed the prominent eastern
shell. The western shell of G150.3+4.5 appears to be fainter. The new
SNR G149.5+3.2 can also be identified in the lower right of
G150.3+4.5.

\citet{Gerbrandt14} demonstrate the non-thermal synchrotron emission
nature of the eastern shell of G150.3+4.5 and propose that it is a new
SNR. We tested it with the TT-plot method \citep{Turtle62}, using the
Urumqi $\lambda$6\ cm, the Effeslberg $\lambda$11\ cm, and the
Effelsberg $\lambda$21\ cm data. We got the brightness temperature
spectral index of $\beta = -2.40\pm0.17$ between the $\lambda$6\ cm
and $\lambda$21\ cm data, and $\beta = -2.38\pm0.42$ between the
$\lambda$6\ cm and $\lambda$11\ cm data, which are consistent with the
result of \citet{Gerbrandt14}. The western shell of G150.3+4.5 is not
mentioned in \citet{Gerbrandt14}. It can be identified from all total
intensity images at various wavelengths as we showed. This shell and
the eastern one both curve to the southern lower section, suggesting
that they are from the same entity. For the emission that is free of
the contamination by the new SNR G149.5+3.2 (see the rectangle area b
as indicated in Fig.~\ref{tp}), we tested the spectrum of the upper
part of the western shell of G150.3+4.5. TT plots give a spectral
index of $\beta = -2.69\pm0.24$ between the Urumqi $\lambda$6\ cm and
the Effelsberg $\lambda$21\ cm data, and $\beta = -2.44\pm0.68$
between the Urumqi $\lambda$6\ cm and Effelsberg $\lambda$11\ cm data,
indicating a non-thermal nature.

\subsection{Other noticeable structures}

Three noticeable structures other than the new SNR G150.3+4.5 are well
resolved in the $\sim$ 1$\arcmin$ resolution CGPS $\lambda$21\ cm
data. They are all located in the lower left hand corner of the maps
in Fig.~\ref{tp} and collected by \citet{Kerton07} in the CGPS
extended source catalog. G151.2+2.6 (CGPSE~169) has a circular shape
with a size of about 40$\arcmin$, while G150.9+2.7 (CGPSE~168) and
G151.4+3.0 (CGPSE~172) both have an arc shape with a length of about
20$\arcmin$. G151.2+2.6, however, is not detected at 408~MHz. Thus its
nature is still unclear. Based on the steep non-thermal spectral
indices found between 408~MHz and 1420~MHz CGPS data ($\alpha =
-1.3\pm0.3$ for G150.9+2.7, $\alpha = -0.4\pm0.2$ for G151.4+3.0),
\citet{Kerton07} propose that the two arcs G150.9+2.7 and G151.4+3.0
form a new Galactic SNR G151.20+2.85.

In the Urumqi and Effelsberg maps with a $\sim$10$\arcmin$ resolution
, the circular G151.2+2.6 can be recognized in all three frequency
maps. G150.9+2.7 is mixed with the upper part of G151.2+2.6 and cannot
be distinguished. Instead of being an arc, G151.4+3.0 is seen to be an
elongated point-like source.

To know their properties, we determined the spectra first. We
subtracted the point-like sources from the Urumqi $\lambda$6\ cm, the
Effelsberg $\lambda$11\ cm and the Effelsberg $\lambda$21\ cm data
based on the NVSS catalog. For the circular G151.2+2.6, to avoid the
contamination from G150.9+2.7, the TT plot is only made for the region
below $b = 2\fdg6$. We obtained the brightness temperature spectral
index of $\beta = -2.03\pm0.32$ between $\lambda$6\ cm and
$\lambda$11\ cm, and $\beta = -2.18\pm0.08$ between $\lambda$6\ cm and
$\lambda$21\ cm. The flat radio continuum spectrum implies that
G151.2+2.6 may have a thermal origin.

For G150.9+2.7, we cannot see it owing to the overlap with
G151.2+2.6. For the source G151.4+3.0, a non-thermal spectral index is
found to be $\beta = -2.62\pm0.15$ between the $\lambda$6\ cm and
$\lambda$21\ cm data, supporting the idea of \citet{Kerton07} that
G151.4+3.0 is a (part of) SNR.

\subsection{Hints in other bands}

We search for coincidence of the objects we discussed in infrared,
H$\alpha$, and DSS2 red maps \citep{Mclean00}. From the ancillary
60$\mu$m infrared map downloaded from the CGPS website, except for the
western part of G151.2+2.6, we do not see any strong correlations
between the infrared and the radio continuum emission toward any
extended objects in Fig.~\ref{tp} (see Fig.~\ref{other}, {\it left
  panel}). \citet{Kerton07} suggest that there is related infrared
emission in the eastern shell of G150.3+4.5. Together with the flat
spectral index they derived, the eastern shell was said to be an
\ion{H}{II} region. We agree with \citet{Gerbrandt14} that the strong
overall infrared emission does not appear to be correlated with the
eastern shell of the new SNR G150.3+4.5.

We extracted the H$\alpha$ image from the Wisconsin H$\alpha$ mapper
northern sky survey \citep{Haffner03}. Again we do not see any firm
correlation with any structures except for some parts of G151.2+2.6
(Fig.~\ref{other}, {\it middle panel}). Strong H$\alpha$ emission is
seen in the eastern part of the image, and it overlaps with the
eastern shell of the new SNR G150.3+4.5. This could possibly explain
the non-detection of the polarized emission at $\lambda$6\ cm as a
result of depolarization, if the magneto-ionized medium appears as the
foreground.

From the DSS2 red image, \citet{Gerbrandt14} show that an optical
filamentary structure is coincident with the lower part of the eastern
shell of G150.3+4.5. For the other parts of G150.3+4.5, we do not see
this coincidence. A ring-shaped optical counterpart is found for
G151.2+2.6 (see Fig.~\ref{other}, {\it right panel}). This, together
with the flat radio continuum spectrum, the H$\alpha$, and infrared
emission, strongly suggest that it is an \ion{H}{II} region.

\section{Summary}

By combining the Urumqi $\lambda$6\ cm Galactic plane survey and the
new medium latitude survey, we discovered an enclosed oval-shaped
object G150.3+4.5. It contains three shells in the east, west, and
north, respectively. Effelsberg $\lambda$11\ cm and $\lambda$21\ cm
radio continuum data are used for the spectral index analysis, and the
high angular resolution CGPS $\lambda$21\ cm and $\lambda$73.5\ cm
data are used for its fine structure. The eastern and western shells
of G150.3+4.5 can be firmly confirmed, while the faint northern shell
seen at $\lambda$6\ cm has not been detected in other radio
frequencies.

The TT plots between $\lambda$6\ cm, $\lambda$11\ cm, and
$\lambda$21\ cm are employed for analyzing the spectra of the eastern
and western shells of the newly discovered loop of G150.3+4.5. We
found non-thermal radio spectra for the shells on both sides. The
spectral index for the eastern shell is about $\beta \sim -2.4$, and
about $\beta \sim -2.7$ for the western shell. From the properties of
this half loop and the optical filamentary structrure that is related
to the eastern radio shell \citep{Gerbrandt14}, G150.3+4.5 should be a
SNR, though the polarized emission from G150.3+4.5 is not detected,
which might be due to the sensitivity limit or depolarization effect
in the eastern shell.

For the other prominent structures in the observed area, we confirm
that the arc structure G151.4+3.0 has a non-thermal spectrum, based on
the Urumqi $\lambda$6\ cm and the Effelsberg $\lambda$21\ cm
data. This agrees with the result of \citet{Kerton07} that it has a
SNR origin. But from the morphology, it is still not clear G150.9+2.7,
and G151.4+3.0 form one SNR or two separated SNRs. We found a flat
radio continuum spectrum for the circular G151.2+2.6 and the related
H$\alpha$ and infrared emission in its western part, and also the
related optical emission in the DSS2 red image, which all suggest that
G151.2+2.6 is an \ion{H}{II} region.

\begin{acknowledgements}
We thank the referee for helpful comments that improved our paper. The
authors are supported by the National Natural Science Foundation of
China (11303035) and by the Strategic Priority Research Program ``The
Emergence of Cosmological Structures'' of the Chinese Academy of
Sciences, Grant No. XDB09010200. X.Y.G is also supported by the Young
Researcher Grant of National Astronomical Observatories, Chinese
Academy of Sciences. We acknowledge Mr Otmar Lochner for construction
of the excellent $\lambda$6\ cm receiver and Mr Maozheng Chen and Jun
Ma for their skillful maintenance.

\end{acknowledgements}

\bibliographystyle{aa}
\bibliography{bbfile}

\end{document}